\documentclass[lettersize,journal]{IEEEtran}
\IEEEoverridecommandlockouts
\usepackage[utf8]{inputenc} 
\usepackage[T1]{fontenc}
\usepackage{url}
\usepackage{ifthen}
\usepackage{cite}
\usepackage[cmex10]{amsmath} 
\usepackage{amssymb}
\usepackage{graphicx}
\usepackage{wrapfig}
\usepackage{multirow}
\usepackage{array}
\usepackage{makecell}
\usepackage{tikz}
\usepackage{algorithm}
\usepackage{algpseudocode}
\usepackage{comment}
\usepackage[english]{babel}

\newtheorem{thm}{Theorem}

\newtheorem{prop}{Proposition}

\newtheorem{defn}{Definition}

\newtheorem{exmp}{Example}

\begin{document}
	
	\title{Secrecy and  Privacy  in Multi-Access Combinatorial Topology}
	
	\author{Mallikharjuna Chinnapadamala and B. Sundar Rajan \\ Department of Electrical Communication Engineering, Indian Institute of Science, Bengaluru \\
		E-mails: chinnapadama@iisc.sc.in, bsrajan@iisc.ac.in}
	
	
	
	\maketitle
\begin{abstract}
In this work, we consider the multi-access combinatorial topology with $C$ caches where each user accesses a unique set of $r$ caches. For this setup, we consider secrecy, where each user should not know anything about the files it did not request, and demand privacy, where each user's demand must be kept private from other non-colluding users. We propose a scheme satisfying both conditions and derive a lower bound based on cut-set arguments. Also, we prove that our scheme is optimal when $r\geq C-1$, and it is order-optimal when the cache memory size $M$ is greater than or equal to a certain threshold for $r<C-1$. When $r=1$, in most of the memory region, our scheme achieves the same rate as the one given by the secretive scheme for the dedicated cache setup by Ravindrakumar et al. ( 'Private Coded Caching,' in \textit{IEEE Transactions on Information Forensics
and Security}, 2018), while satisfying both secrecy and demand privacy conditions.
\end{abstract}
\section{Introduction}
Coded caching is an effective technique for alleviating network congestion during peak traffic hours by storing portions of content in users' local caches. The model consists of a single server that serves multiple users through a shared, error-free link. The concept of coded caching was initially introduced by Maddah-Ali and Niesen (MAN) \cite{MaN}, where each user possesses a dedicated cache. The coded caching scheme operates in two phases: the placement phase, during which each user's cache is populated up to its capacity, and the delivery phase, where users declare their demands, and the server responds accordingly. The server leverages the cached content in this phase to minimize network traffic. It has been demonstrated that a "global caching gain" can be achieved by serving multiple users simultaneously with a single transmission. The number of files the server sends to satisfy the demands of all the users is called the rate of the system.

Multi-access networks, where each user has access to multiple caches, have been studied in \cite{MKR}, \cite{RaK}, \cite{SPE}, \cite{HKD}, and \cite{KMR}. In \cite{MKR}, each user is connected to a distinct set of $r$  caches out of a total of $C$  caches, ensuring that for every set of $r$ caches, there is a corresponding user. This network structure is known as combinatorial topology \cite{BrE}. The scheme proposed in \cite{MKR} was proven to be optimal under the assumption of uncoded placement in \cite{BrE}. A more generalized version of the combinatorial topology was introduced in \cite{BrE}, where different users can be connected to varying numbers $ r \in [0:C]$ of caches, and each unique set of $r$ caches is assigned to $ K_r $ users, maintaining this property for every $ r \in [0:C] $. Additionally, a multi-access network comprising $ K $ users and $K$  caches, where each user has access to $r$ caches in a cyclic wrap-around manner, was analyzed in \cite{RaK}, \cite{SPE}, \cite{HKD}, and \cite{KMR}.

For the dedicated cache setup in \cite{MaN}, an additional constraint was introduced in \cite{RPKP}, ensuring that no user gains any information about the content of files other than the one it requested, whether from its cache or the server's transmissions. This was referred to as private coded caching in \cite{RPKP} and secretive coded caching in \cite{RPKP2}. We use the latter terminology. The secretive coded caching was extended to other settings that include shared cache networks \cite{MeR}, \cite{PNR}, device-to-device networks \cite{ZY}, combination networks \cite{ZY2}, and fog radio access networks \cite{TJHZN}. The secretive coded caching for dedicated cache setup when users are colluded is considered in \cite{MaS}. However, to the best of our knowledge, secretive coded caching has not been explored in a multi-access combinatorial topology.
\begin{figure} 
     \centering
     \includegraphics[width=\linewidth]{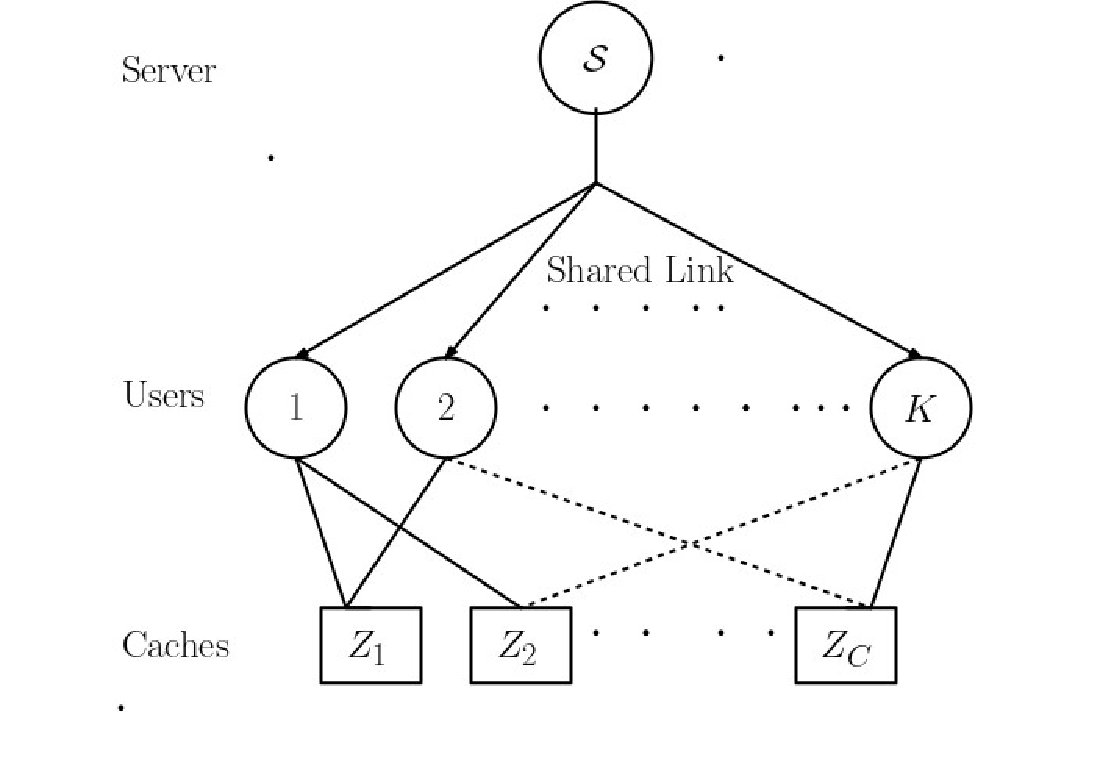}
     \caption{Combinatorial topology.}
     \label{Fig1}
\end{figure}

The concept of demand privacy, which ensures that no user can infer the file index requested by another user, has been studied in \cite{WaC}, \cite{Kam}, \cite{AST}, and \cite{YaT2}. Content security, which aims to protect library content from an eavesdropper who intercepts the server's transmission during the delivery phase, was examined in \cite{SeT}. The feasibility of simultaneously achieving both demand privacy and security using the key superposition approach was demonstrated in \cite{YaT}. Additionally, demand privacy for multi-access networks, where users access $r$ caches in a cyclic wrap-around manner \cite{HKD}, was investigated in \cite{NaR} and \cite{LWCC}. Security and demand privacy were considered simultaneously for multi-access combinatorial topology in \cite{ChR}. Security and demand privacy for combinatorial topology with private cache was studied in \cite{ChR2}. However, to the best of our knowledge, secrecy and demand privacy simultaneously has not been considered in multi-access combinatorial topology. In this work, we consider secrecy and demand privacy simultaneously for combinatorial topology. 
\subsection{System Model}
We consider the combinatorial topology as shown in Fig. \ref{Fig1}. The model consists of a server with $N$ files connected to $K$ users through an error-free shared link. There are $C$ caches. Each user has access to a unique set of $r$ out of $C$ caches. For every set of $r$ caches, there is a user. So, total number of users is $K=\binom{C}{r}$. The size of each cache is $M$ files. The content of the cache $c\in[C]$ is represented as $Z_{c}$.
\subsection{Our Contributions}
We consider secrecy and demand privacy simultaneously for combinatorial topology. The following are the technical contributions:
\begin{itemize}
    \item A secretive and demand private coded caching scheme that simultaneously provides secrecy and privacy is proposed for combinatorial topology.
    \item A lower bound based on the cut-set arguments is derived. The proposed scheme is proved to be optimal when $r\geq C-1$ for $N\geq2K$.
    \item For $r < C-1$, the proposed scheme is proved to be within a multiplicative gap of $5$ from the optimal when $M\geq\frac{N}{C-r}+\binom{C-1}{r-1}$ and $K\leq5(r+1)$.
    \item For the same memory-rate pair achieved by the secretive scheme in \cite{RPKP}, the proposed scheme achieves both secrecy and privacy when $r=1$, except in the small memory region from  $M=\frac{N(C-2)}{2}+1$ to $M=N(C-1)+1$. 
    \item We numerically compare our proposed scheme with the secretive scheme for dedicated cache setup in \cite{RPKP} in the following ways:
    \begin{itemize}
        \item Firstly, we consider the total number of caches in both the dedicated cache and multi-access setup to be the same. The size of each cache is also the same. But, the multi-access setup supports more number of users ($K=\binom{C}{r}$) than the dedicated cache setup for a given $r$. So, we compare the Rate Per User (RPU), which is the rate normalized with the number of users instead of the rate. The total system memory is also the same in both the dedicated cache and the multi-access setup. In this setting, we see that the multi-access setup performs better than the dedicated cache setup.
        \item Then, we consider the case where the total number of caches is the same, and each user accesses the same amount of memory. To make the total memory accessed per user the same for both the dedicated cache and the multi-access settings, we make the cache size of the dedicated cache setup $r$ times more than the cache size of the multi-access setup. The multi-access setup is at a disadvantage here as the total memory of the system is $r$ times more for the dedicated cache setup than the multi-access setup. In this setting, we see that the dedicated cache performs better than the multi-access setup.
    \end{itemize}
\end{itemize}
\subsection{Paper Organisation}
Section \ref{sec2} introduces problem setup and preliminaries. Section \ref{sec3} contains the main results. Section \ref{sec4} gives the proof of Theorem \ref{th1} and Section \ref{sec5} contains the proofs of Theorem \ref{th2} and Theorem \ref{th3}. Section \ref{sec6} contains numerical evaluations, and Section \ref{sec7} concludes the paper. \\

 \textit{Notations:} The set $\{1,2,3,....n\}$ is denoted as $[n]$. $|\mathcal{X}|$ denotes cardinality of the set $\mathcal{X}$.  For two non-negative integers $l,m$ such that $l\leq m$, we use $[l:m]$ to denote the set $\{l,l+1,....m\}$.. $\lfloor x\rfloor$ gives the largest integer lesser than or equal to $x$. A finite field of size $q$ is represented as $\mathcal{F}_{q}$. For a set $\mathcal{A}$, we define $X_{\mathcal{A}} := \{X_{i} : i \in \mathcal{A}\}$.  For two sets $\mathcal{A}$ and $\mathcal{B}$, $\mathcal{A} \backslash \mathcal{B}$ denotes the elements in $\mathcal{A}$ but not in $\mathcal{B}$ and $X_{\mathcal{A}, \mathcal{B}}:=\{X_{i,j} : i \in \mathcal{A}, j \in \mathcal{B}\}$. For two non-negative integers $n,m$, we have $\binom{n}{m} = \frac{n!}{(n-m)!m!}$, if $n\geq m$, and $\binom{n}{m} =0$ if $n<m$. For a matrix $\textbf{D}$, $[\textbf{D}]_{ij}$ represents the $(i,j)^{th}$ entry of $\textbf{D}$. The standard unit vector of length $N$, with one in the $n^{th}$ position
and zero elsewhere is denoted by $\textbf{e}_{n}\in \mathcal{F}_{2}^{N}$.
\section{Problem Setup and Preliminaries}\label{sec2}
In this section, we introduce the problem setup first and then discuss some preliminaries.

The problem setup is shown in Fig.\ref{Fig1}. The server $\mathcal{S}$ consists of a library of $N$ files, $W_{[N]}= \{W_{1}, W_{2},....W_{N}\}$ each of size $F$ bits. Let $\mathcal{K}$ denote the set of $K$ users that are connected to the server through an error-free shared link. There are $C$ caches, each of size $M$ files. Each user is connected to a unique set of $r$ caches out of $C$ caches. Let,
 \begin{equation} \label{eq1}
 	{\Omega_{r} \triangleq \{\mathcal{G} : \mathcal{G} \subseteq [C],|\mathcal{G}|=r\}}.
 \end{equation}
 For every set of $r$  caches, there is a user. So, we have  $K=\binom{C}{r}$ users and these are represented as $U_{\mathcal{G}}, \mathcal{G} \in \Omega_{r}$.    

 
 \textit{Placement Phase}: During this phase, the server fills the caches. The server privately generates a randomness $P$ from  some finite alphabet $\mathcal{P}$ and fills the  caches of each user by using the cache function
	\begin{equation} \label{eq2}
		{\psi_{c} :\mathcal{P} \times (\mathcal{F}_2)^{NF} \rightarrow (\mathcal{F}_2)^{MF}, c \in [C]}. 
	\end{equation}
 The cached content of the cache is denoted by,
	\begin{equation} \label{eq3}
		{Z_{c}:= \psi_{c}(P,W_{[N]})}. 
	\end{equation}
     
 A user $U_{\mathcal{G}}$ has access to  cache content $Z_{c}$ if $c\in \mathcal{G}$. $Z_{\mathcal{G}}$ represents the total cache content available to the user $U_{\mathcal{G}},$ i.e.,
	\begin{equation*} 
		{Z_{\mathcal{G}}= \cup_{i\in \mathcal{G}} Z_{i}}.
	\end{equation*} 
 
The placement is done in such a way that no user knows anything about any file from the contents of the caches accessible to it. i.e.,
\begin{equation}\label{eq4}
    I(W_{[N]}; Z_{\mathcal{G}})=0, \forall \mathcal{G} \in \Omega_{r}.
\end{equation}
 \textit{Delivery Phase}: The demand of a user $U_{\mathcal{G}}$ is ${d}_{\mathcal{G}}$, i.e., each user wants the file $W_{d_{\mathcal{G}}}$. As each user is associated with a $r$-sized subset, arrange all the users in the lexicographic order of their associated subsets. Let $\textbf{d}$ be the demand vector containing the demands of all the users in that order.
 In order to satisfy the demand of every user, the server transmits 
	\begin{equation} \label{eq5}
		{X:=\phi(P,\textbf{d},W_{[N]})},
	\end{equation}
	where ${\phi :\mathcal{P}\times[N]^{K} \times(\mathcal{F}_2)^{NF} \rightarrow (\mathcal{F}_2)^{RF}}$ and the quantity $R$ is called the rate of the system. The file library $W_{[N]}$, the randomness $P$ and the demands \{${{d}_{{\mathcal{G}}}} : \mathcal{G} \in \Omega_{r} $ \} are independent. The following conditions are required to be satisfied by the delivery scheme:
 \begin{itemize}
		\item \textit{Correctness}: Each user should be able to retrieve its demanded file,
		\begin{equation} \label{eq6}
			H(W_{{d}_{{\mathcal{G}}}}|X,{{d}_{{\mathcal{G}}}},Z_{{\mathcal{G}}}) = 0,\quad  \forall \mathcal{G}\in\Omega_{r}.
		\end{equation}
		\item \textit{Secrecy}: No user should know anything about the content of the files it did not request, either from its caches or from the server transmissions,
		\begin{equation} \label{eq7}
			I(W_{[N] \backslash d_{\mathcal{G}}};X,Z_{\mathcal{G}}) = 0.
		\end{equation}
		\item \textit{Privacy}: Any non-colluding user should not know anything about the demands of the other users. Let $\textbf{d}_{\bar{\mathcal{G}}}$ represent the demands of all users except $U_{\mathcal{G}}$,
		\begin{equation} \label{eq8}
			I(\textbf{d}_{\bar{\mathcal{G}}};X, {d}_{{\mathcal{G}}},Z_{{\mathcal{G}}}) = 0 \quad  \forall \mathcal{G} \in \Omega_{r}.
		\end{equation}
	\end{itemize}
    \begin{defn} \label{def1}
         We say that the pair $(M, R)$ is achievable if there exists a scheme that satisfies the above three conditions with rate $R$ and memory $M$. We refer to the scheme as a secretive and private scheme. The optimal rate for the given setting is defined as:
 \begin{equation} \label{eq9}
     R^{*}= \inf\{R : (M, R) \quad \text{is achievable} \}.
 \end{equation}
    \end{defn}
In the following subsections, we discuss some preliminaries.
\subsection{Secret Sharing schemes} The coded caching schemes that studied secrecy in the literature used non-perfect secret sharing schemes. The key idea is to encode the secret into shares in such a way that accessing a certain number of shares does not reveal any information about the secret, and accessing all the shares enables the recovery of the secret completely. The formal definition of a non-perfect secret-sharing scheme is given below.
    \begin{defn}(\cite{CDN})
        For a secret $W$ with size $B$ bits and $m<n$, an $(m,n)$ non-perfect secret sharing scheme generates $n$ equal-sized shares $S_{1},S_{2},..,S_{n}$ such that accessing any $m$ shares does not reveal any information about the secret $W$ and $W$ can be completely reconstructed from all the $n$ shares, i.e,
        \begin{subequations}
        \begin{align}\label{eq10}
            &I(W;S)=0, \forall S \subseteq \{S_{1},S_{2},...S_{n} \}, \text{such that} |S|\leq m, \\
            &H(W|S_{1},S_{2},...S_{n})=0.
            \end{align}
        \end{subequations}
    \end{defn}
    The size of each share should be at least $\frac{B}{n-m}$ bits in a non-perfect secret sharing scheme\cite{RPKP}. For large enough $B$, there exists non-perfect secret sharing schemes with the size of each share equal to $\frac{B}{n-m}$ bits.\\
    \subsection{Cauchy matrix}
    \begin{defn}(\cite{PlX})
        Let $\mathcal{F}_{q}$ be a finite field, with $q\geq u+v$. Let $X=\{x_{1}, x_{2},...,x_{u} \}$ and $Y=\{y_{1}, y_{2},...,y_{v} \}$ be two distinct sets of elements in $\mathcal{F}_{q}$ with $X \cap Y = \emptyset$ such that: \\
        i. $ \forall i\in [u], \forall j\in [v]: x_{i}-y_{j}\neq0.$\\
        ii. $\forall i, j \in [u]: i \neq j, x_{i} \neq x_{j} $ and $\forall i, j \in [v], i \neq j, y_{i} \neq y_{j},$ \\
        where all operations are performed over $\mathcal{F}_{q}$. Then, the matrix,
        \[
A =
\begin{bmatrix}
\frac{1}{x_1 - y_1} & \frac{1}{x_1 - y_2} & \cdots & \frac{1}{x_1 - y_v} \\
\frac{1}{x_2 - y_1} & \frac{1}{x_2 - y_2} & \cdots & \frac{1}{x_2 - y_v} \\
\vdots & \vdots & \ddots & \vdots \\
\frac{1}{x_u - y_1} & \frac{1}{x_u - y_2} & \cdots & \frac{1}{x_u - y_v}
\end{bmatrix}
\]
is a Cauchy matrix of dimension $u\times v$. A Cauchy matrix has full rank, and each of its sub-matrices is also a Cauchy matrix.
         \end{defn}
\subsection{Secretive Coded Caching Scheme in \cite{RPKP}  }
    The secretive coded caching scheme for dedicated cache setup \cite{MaN} is given in \cite{RPKP}. Each file is encoded using a $(\binom{K-1}{t-1}, \binom{K}{t})$ non-perfect secret sharing scheme \cite{CDN}, where $t=\frac{K(M-1)}{N+M-1}\in \{0,1,2..K-2\}$. When $M=N(K-1)$, the scheme encodes each file with a $(K-1,K)$ secret sharing scheme. The server places the shares along with randomly generated keys of length the same as that of a share in each user's cache during the placement phase. Once all users reveal their demands to the server, it sends $\binom{K}{t+1}$ transmissions over the shared link. Each transmission is an XOR'ed combination of shares, further XOR'ed with a key. The secretively achievable rate $R(M)$ is given by :
    \[
    R(M) = \begin{cases}
        \frac{K(N+M-1)}{N+(K+1)(M-1)} \quad\text{for} \hspace{0.1cm} M=\frac{Nt}{K-t}+1, \\
        1, \quad \text{for} \hspace{0.1cm}   M=N(K-1),
    \end{cases}
    \]
    where $t\in \{0,1,2,..K-2\}$.
    \subsection{Multi-access Coded Caching\cite{MKR}}
A multi-access network consisting of $C$ caches and \( K \) users, with each user having access to a unique set of \( r \) caches, was considered in \cite{MKR}. For every set of \( r \) caches, there is a user. This network can accommodate a large number of users at low subpacketization levels. This network is referred to as combinatorial topology \cite{BrE}, and the scheme given in \cite{MKR} was proved to be optimal under the assumption of uncoded placement in \cite{BrE}. The well-known Maddah-Ali–Niesen (MAN) scheme becomes a special case of the scheme proposed in \cite{MKR} when \( r=1 \).  

The problem setup is as follows: The server containing \( N \) files, denoted as \( W_{1}, W_{2}, W_{3}, \dots, W_{N} \), each of size \( F \) bits, is connected to \( K \) users through an error-free shared link, with each user having access to a unique set of \( r \) out of \( C \) caches. Each cache has a size of \( M \) files, and \( Z_{c} \) denotes the content of the cache \( c \in [C] \). The total number of users, $K=\binom{C}{r}$. During the placement phase, the server divides each file into \( \binom{C}{t} \) subfiles, where \( t = \frac{CM}{N} \), and the content of the cache \( c \) is given by,  

\begin{center}  
    \( Z_{c} = \{W_{i,\mathcal{T}} : c\in\mathcal{T}, \mathcal{T} \subset C, |\mathcal{T}| =t \quad \forall \quad i \in [N] \} \).  
\end{center}  

Let \( d_{U} \), where \( U \subset C \), \( |U|=r \), denote the demand of the user connected to the set of caches represented by \( U \). Then, for each \( S \subset C \), \( |S|=t+r \), the server transmits  

\begin{center}  
    \( \underset{\substack{U\subset S\\|U|=r}}{\bigoplus} W_{{{d_{U}}},S\backslash U}. \)  
\end{center}  

Thus, the scheme achieves the rate given by:  

\begin{center}  
    \( R=\frac{\binom{C}{t+r}}{\binom{C}{t}}. \)  
\end{center}  


\section{Main Results} \label{sec3}
In this section, we present the main results of this paper. Theorem \ref{th1} presents a lower bound on the optimal rate described in Definition \ref{def1}. Theorem \ref{th2} presents a scheme that achieves both secrecy and privacy. Theorem \ref{th3} presents a scheme that achieves only secrecy. Finally, Theorem \ref{th4} presents optimality results.
\begin{thm}\label{th1}
    For the multiaccess combinatorial topology described in Section \ref{sec2}, when $M\geq \frac{K-1}{C-r}$, the optimal rate is lower bounded as
    \begin{multline*}         
    R^{*}(M) \geq \max_{l \in \{1,2,...\min(N/2,K)\}} \frac{l \lfloor N/l \rfloor-1-(z-r) M)}{\lfloor N/l \rfloor - 1}, 
    \end{multline*} 
    where $z=(\min(l+r-1,C)$.
\end{thm}
 \textit{Proof}: Refer to Section \ref{sec4} for the proof.
 
 By substituting $l=1$ in the above inequality, we get $R^{*}(M)\geq 1$. So, this is the minimum achievable rate to satisfy the secrecy condition. This implies that the user cannot know anything about the requested file from the cached content alone and needs $F$ bits from the server to learn anything about it.
 \begin{thm}\label{th2}
     Consider the combinatorial topology described in Section \ref{sec2}. For $M=\frac{N\binom{C-1}{t-1}}{\binom{C-r}{t}}+\binom{C-1}{r-1}$ with $t\in[0:C-r]$, there exists a scheme that satisfies the conditions in (\ref{eq6}), (\ref{eq7}), and (\ref{eq8}) with the rate
     \begin{equation*}
         R=\frac{\binom{C}{t+r}}{\binom{C-r}{t}}.
     \end{equation*}
 \end{thm}
 \textit{Proof:} The scheme that achieves the above rate is presented in Section \ref{ssec5a}.
 
  When $r=1$, for the same cache memory and rate in \cite{RPKP}, the proposed scheme provides both secrecy and privacy except in the small memory region from  $M=\frac{N(C-2)}{2}+1$ to $M=N(C-1)+1$. The scheme in \cite{RPKP} provides only secrecy. Our scheme is defined for the values of $M$ that result from the integer values of $t$. For other values of $M$, memory sharing is used to obtain the rate. 
 \begin{thm} \label{th3}
 Consider the combinatorial topology described in Section \ref{sec2}. For $M=\frac{N\binom{C-1}{t-1}}{\binom{C-r}{t}}+\binom{C-1}{r-1}$ with $t\in[0:C-r-1]$, there exists a scheme that satisfies the conditions in (\ref{eq6}), and (\ref{eq7}) with the rate
     \begin{equation*}
         R=\frac{\binom{C}{t+r}}{\binom{C-r}{t}}.
     \end{equation*}
     For $M=N\binom{C-1}{r}$, the rate $R=1$ is achievable.
 \end{thm}
 \textit{Proof:} The scheme that achieves the above rate is described in Section\ref{ssec5}.
 
 We refer to this scheme as the secretive scheme. When $r=1$, our setup reduces to a dedicated cache setup, and the scheme reduces to the scheme given in \cite{RPKP}.
 \begin{thm}\label{th4}
      For the multi-access combinatorial topology described in Section \ref{sec2}, the achievable rate $R$ of the scheme that satisfies both secrecy and privacy conditions and the optimal rate $R^{*}$ satisfy \\
    \[
    \frac{R(M)}{R^{*}(M)} \leq \begin{cases}
        1, \quad r \geq C-1, N\geq2K .\\
        5, \quad   M\geq \frac{N}{C-r}+\binom{C-1}{r-1}, r < C-1,\\ \hspace{1cm} K\leq 5(r+1).
            \end{cases}
    \]
 \end{thm}
 \textit{Proof}: Refer the Appendix.
 \section{Proof of Theorem \ref{th1}} \label{sec4}
 In this section, we derive a cut-set based lower bound on $R^{*}(M)$ given in Theorem \ref{th1}. First, we prove that the minimum cache memory required to satisfy the secrecy condition is $M\geq \frac{K-1}{C-r}$.
 \begin{prop} \label{prop1}
For a multi-access combinatorial topology, the system satisfies the secrecy condition in (\ref{eq7}) when $ N \geq K $, if the cache memory size $ M $ satisfies the following inequality:
\begin{equation} \label{eq11}
    M \geq \frac{K-1}{C-r}.
\end{equation}
\end{prop}
	\textit{Proof}: Consider the system model shown in Fig.\ref{Fig1}. The server containing $N$ files is connected to $K$ users through an error-free shared link. Each user can access a unique set of $r$ caches out of $C$  caches, each of size  $ M$. For the considered network, the total number of users is $K=\binom{C}{r}$. So, each $r$ sized subset defines a user. As each user is indexed by an $r$ sized subset, let's arrange all $\binom{C}{r}$ users in lexicographic order. Let $i^{th}$ user demand the file $W_{d_{i}}$ and $\tilde{Z}_{i}$ represent the total cache content available to the $i^{th}$ user. Let, \\
    \begin{equation*}
    {W}= \{ W_{d_{1}}, W_{d_{2}},....W_{d_{K}}\}, \quad \quad \quad
    {Z}= \{ Z_{1}, Z_{2},....Z_{C}\}. 
    \end{equation*}
    By correctness condition in (\ref{eq6}) and the secrecy condition in(\ref{eq7}), we can write
		\begin{equation}\label{eq12}
		H(W_{d_{1}},W_{d_{2}},.....W_{d_{K}}|X,{Z})=0,
		\end{equation}
		\begin{equation}\label{eq13}
		I(W_{d_{1}}, W_{d_{2}},..,W_{d_{k-1}}, W_{d_{k+1}}...W_{d_{K}}; X,\tilde{Z}_{k})=0, \forall k \in [K].
		\end{equation}
		 We have,
		\begin{subequations}
			\begin{align}
			    \binom{C}{r}F& = H({W}) \\
			&=I({W};X,{Z})  +H({W}|X,{Z}) \\
			&= I({W};X,{Z}) \label{eq14c}\\
            &= I({W};X,\tilde{Z}_{k})+I({W}; {Z} \backslash \tilde{Z}_{k} |X,\tilde{Z}_{k}) \\
			&=I({W}\backslash W_{d_{k}};X,\tilde{Z}_{k})+I(W_{d_{k}}; X, \tilde{Z}_{k}| {W}\backslash W_{d_{k}})+ \\ \notag
            & \quad I({W};{Z}\backslash \tilde{Z}_{k}|X,\tilde{Z}_{k}) \\
            &= I(W_{d_{k}}; X,\tilde{Z}_{k}| {W}\backslash W_{d_{k}})+I({W};{Z}\backslash \tilde{Z}_{k}|X,\tilde{Z}_{k}) \label{eq14f} \\
            &\leq H(W_{d_{k}})+ H({Z}\backslash \tilde{Z}_{k}) \\
			&\leq F+(C-r)MF.
		\end{align}
		\end{subequations}
        
		This implies $M \geq \frac{K-1}{C-r}$, where  (\ref{eq14c}) and (\ref{eq14f}) come from correctness in (\ref{eq12}) and secrecy conditions in (\ref{eq13}) respectively.\\
        This completes the proof of Proposition \ref{prop1}.

\textit{Proof of Theorem \ref{th1}}: Now, we prove the lower bound on $R^{*}(M)$.
  Consider the combinatorial topology described in Section \ref{sec2}. Each user is connected to a unique set of $r$ caches out of $C$ caches, each of size $M$. The content of a cache is represented as $Z_{i},i \in [C]$. As each user is associated with an $r$ sized subset, arrange all $\binom{C}{r}$ users in the lexicographic order of their associated subsets. Each user is connected to $r$ out of $C$  caches. So, if the first $l$ users are considered, each user is connected to $r$ out of $z=\min(l+r-1, C)$ caches. The server consists of a library of $N$ files, $W_{[N]}=\{W_{1}, W_{2},.....W_{N}\}$ each of size $F$ bits. Consider the first $l$ users,  let them request $W_{1},....W_{l}$, respectively. Let the server make the transmission $X_{1}$. Using the transmission $X_{1}$, and the content of the first $z=\min(l+r-1,C)$ caches, the first $l$ users can decode $W_{1},....W_{l}$. Now, let the first $l$ users request $W_{l+1}, W_{l+2},....W_{2l}$, respectively. The server sends the transmission $X_{2}$. Using the transmission $X_{2}$, and the first $z$ caches, the first $l$ users can decode the files $W_{l+1}, W_{l+2},....W_{2l}$. By continuing in this manner, consider the transmission $X_{\lfloor N/l \rfloor}$. The first $l$ users can decode $W_{(\lfloor N/l \rfloor-1)l+1}, W_{(\lfloor N/l \rfloor-1)l+2},.....W_{(\lfloor N/l \rfloor)l}$ files using the first $z$  caches. Considering all transmissions $X_{1}, X_{2},..X_{\lfloor N/l \rfloor}$, using the first $z$ caches, the users must be able to decode $W_{1}, W_{2},....W_{l \lfloor N/l \rfloor }$. Consider the $j^{th}$ transmission. Let the $k^{th}$ user demand $W_{d^{j}_{k}} := W_{(j-1)l+k}$. Let $\tilde{Z}_{k}$ denote the overall content accessible to the $k^{th}$ user. Let,
\begin{align*}
\tilde{W} &= \{ W_{1}, W_{2}, \ldots, W_{l \lfloor N/l \rfloor} \} \setminus \{W_{d^{j}_{k}}\}, \\
\tilde{X} &= \{ X_{1}, X_{2}, \ldots, X_{\lfloor N/l \rfloor} \}, \\
\tilde{X}/s &= \{ X_{1}, X_{2}, \ldots, X_{s-1}, X_{s+1}, \ldots, X_{\lfloor N/l \rfloor} \}, \\
\tilde{Z} &= \{ Z_{1}, Z_{2}, \ldots, Z_{z} \}.
\end{align*}
Along with correctness, we incorporate the condition of secrecy. By the conditions in (\ref{eq6}) and (\ref{eq7}), we get 
 \begin{equation} \label{eq15}
     H(\tilde{W}/\tilde{X},\tilde{Z})=0,
 \end{equation}
 \begin{equation} \label{eq16}
     I(\tilde{W};X_{j}, \tilde{Z}_{k})=0, \quad j=1,2....\lfloor N/l \rfloor,  k\in [l].
 \end{equation}
 
 Now, we have
 \begin{subequations}
     \begin{align}
         (l\lfloor N/l \rfloor-1)F &= H(\tilde{W}) \\
         &= I(\tilde{W};\tilde{X},\tilde{Z})+H(\tilde{W}/\tilde{X},\tilde{Z}) \\
         &= I(\tilde{W};\tilde{X},\tilde{Z}) \label{eq17c} \\
          &= I(\tilde{W}; \{ X_{1}, X_{2},....X_{\lfloor N/l \rfloor} \}, \tilde{Z}) \\
           &= I(\tilde{W}; X_{j},\tilde{Z}_{k})+I(\tilde{W}; \tilde{X}/j, \tilde{Z}\backslash\tilde{Z}_{k}/X_{j},\tilde{Z}_{k}) \\
           &= I(\tilde{W}; \tilde{X}/j, \tilde{Z}\backslash\tilde{Z}_{k}/X_{j},\tilde{Z}_{k}) \label{eq17f} \\
           &\leq H(\tilde{W}; \tilde{X}/j, \tilde{Z}\backslash\tilde{Z}_{k}) \\
           & \leq \sum_{i=1,i\neq j}^{\lfloor N/l \rfloor} H(X_{i})+ H(\tilde{Z}\backslash\tilde{Z}_{k}) \\
           & \leq (\lfloor N/l \rfloor-1) R^{*}F + (z-r)MF, 
         \end{align}
 \end{subequations}
where (\ref{eq17c}) comes from the condition in (\ref{eq15}) and (\ref{eq17f}) comes from the secrecy condition in (\ref{eq16}). By rearranging the terms, we get
  \begin{equation*}
      R^{*} \geq \frac{l \lfloor N/l \rfloor-1-((z-r) M)}{\lfloor N/l \rfloor - 1}.
  \end{equation*}
  optimizing over all possible values of $l$, we obtain
  \begin{equation*}
      R^{*} \geq \max_{l \in \{1,2,...\min(N/2,K)\}} \frac{l \lfloor N/l \rfloor-1-((z-r) M)}{\lfloor N/l \rfloor - 1}.
  \end{equation*}
This completes the proof of Theorem \ref{th1}.

\section{Proofs of Theorem \ref{th2} and Theorem \ref{th3}} \label{sec5}
In this section, we first present a procedure to obtain a secretive and private scheme for combinatorial topology that achieves the rate given in Theorem \ref{th2}. Then, we present a secretive scheme that achieves the rate given in Theorem \ref{th3}.
\subsection{Proof of Theorem \ref{th2}} \label{ssec5a}
In this subsection, we prove Theorem \ref{th2}. The procedure to obtain a secretive and private scheme is described below. 

\textit{Placement Phase}:
For $t\in[0:C-r]$, the first step is to encode each file using a $(\sum_{j=1}^{r} (-1)^{j+1} \binom{r}{j}\binom{C-j}{t-j}, \binom{C}{t})$ non-perfect secret sharing scheme\cite{CDN}. To do this, each file $W_{i}, i \in [N]$ is divided into $\binom{C}{t}-\sum_{j=1}^{r} (-1)^{j+1} \binom{r}{j}\binom{C-j}{t-j}=\binom{C-r}{t}$ equal-sized parts. Let $m=\sum_{j=1}^{r} (-1)^{j+1} \binom{r}{j}\binom{C-j}{t-j}, n=\binom{C}{t}$. So,
\begin{equation} \label{eq18}
    W_{i}=\{W_{i,h}: h\in [n-m]\}, \forall i \in [N].
\end{equation}
Now generate $m=\sum_{j=1}^{r} (-1)^{j+1} \binom{r}{j}\binom{C-j}{t-j}$ encryption keys for each $i\in [N]$ independently and uniformly from $\mathcal{F}_{2}^{F/\binom{C-r}{t}}$. Let those keys be $\{Y_{i,p}: i\in[N], p\in [m]\}$. Consider a $\binom{C}{t}\times \binom{C}{t}$ cauchy matrix\cite{PlX}, $\textbf{C}$. Let $[\textbf{C}]_{uv}=c_{u,v} \forall u,v \in [\binom{C}{t}]$ over $\mathcal{F}_{2^{l}}$, where $2^{l} \geq 2\binom{C}{t}$. The sub-matrices of a Cauchy matrix has full rank. For each file $W_{i}, i \in [N]$, the shares $\{\tilde{W}_{i,\mathcal{T}_f}: f \in [\binom{C}{t}], \mathcal{T}_f \in \Omega_{t}\}$ are generated as follows:

\[
\begin{bmatrix}
    \tilde{W}_{i,\mathcal{T}_{1}} \\
    \tilde{W}_{i,\mathcal{T}_{2}} \\
    \vdots\\
    \tilde{W}_{i,\mathcal{T}_{\binom{C}{t}}}
\end{bmatrix} 
=
\begin{bmatrix}
    c_{1,1} &c_{1,2} &\cdots &c_{1,n} \\
    c_{2,1} &c_{2,2} &\cdots &c_{2,n} \\
    \vdots &\vdots &\ddots &\vdots \\
    
    c_{n,1} &c_{n,2} &\cdots & c_{n,n}
    
\end{bmatrix}
\cdot
\begin{bmatrix}
    W_{i,1} \\
    . \\
    W_{i,n-m} \\
    Y_{i,1} \\
    . \\
    Y{i,m}
\end{bmatrix}
\]
 To protect the server transmissions, the random keys ${\{V_{\mathcal{S}} : \mathcal{S} \in \Omega_{t+r}\}}$ are generated independently and uniformly from $\mathcal{F}_{2}^{F/\binom{C-r}{t}}$. As there are $\binom{C}{r}$ users, the server generates $\binom{C}{r}$ random vectors $\{\textbf{p}_{\mathcal{G}} : \mathcal{G} \in \Omega_{r}\}$ as follows
 \begin{equation} \label{eq19}
\textbf{p}_{\mathcal{G}} \triangleq (p_{\mathcal{G},1},...p_{\mathcal{G},N})^{T} \sim Unif \{\mathcal{F}_{2}^{N}\}, \forall \mathcal{G}\in \Omega_{r}.
\end{equation}
By using the above random vectors, $ \binom{C}{r} \binom{C-r}{t}$ privacy keys, denoted by $\{\tilde{W}_{\textbf{p}_{\mathcal{G}},\mathcal{T}}: \mathcal{G} \in \Omega_{r} , \mathcal{T} \in \Omega_{t},\mathcal{G} \cap \mathcal{T}= \emptyset\}$ are generated as :
		\begin{equation} \label{eq20}
			{\tilde{W}_{\textbf{p}_{\mathcal{G}},\mathcal{T}}:=\underset{n\in[N]}{\bigoplus} p_{\mathcal{G},n}.\tilde{W}_{n,\mathcal{T}}}.
		\end{equation}		
Let us define $D_{\mathcal{G},\mathcal{T}}$ as follows:
	\begin{equation} \label{eq21}
		D_{\mathcal{G},\mathcal{T}} \triangleq \tilde{W}_{\textbf{p}_{\mathcal{G}},\mathcal{T}} \oplus V_{\mathcal{G} \cup \mathcal{T}} ,\quad  \mathcal{G} \in \Omega_{r} , \mathcal{T} \in \Omega_{t},\mathcal{G} \cap \mathcal{T}= \emptyset .
	\end{equation} 
For every $D_{\mathcal{G},\mathcal{T}}$, the following is done:
\begin{itemize}
\item Let $\mathcal{E}$ be any $r-1$ sized subset of $\mathcal{G}$.
    \item Generate $r-1$ random vectors $D^{a}_{\mathcal{G},\mathcal{T}}, a\in\mathcal{E}$ independently and uniformly from $\mathcal{F}_{2}^{F/\binom{C-r}{t}}$.
    \item Now, generate $D^{b}_{\mathcal{G},\mathcal{T}}, b= \mathcal{G} \backslash \mathcal{E}$ as follows 
    \begin{equation} \label{eq22}
        D^{b}_{\mathcal{G},\mathcal{T}} = \bigoplus_{a \in \mathcal{E}} D^{a}_{\mathcal{G},\mathcal{T}}  \oplus D_{\mathcal{G},\mathcal{T}}.
    \end{equation}
\end{itemize}

 The placement of the caches $Z_{c}, c \in [C]$ is done as follows:
 \begin{multline} \label{eq23}
		Z_{c} = \{\tilde{W}_{i,\mathcal{T}} : c\in\mathcal{T} , \mathcal{T} \in \Omega_{t} \quad \forall \quad i \in [N] \} \\ 
		\cup  \quad  \{D^{a}_{\mathcal{G},\mathcal{T}} : c \in \mathcal{G}, \mathcal{G} \in \Omega_{r} ,        \mathcal{T} \in \Omega_{t},  \mathcal{G}\cap \mathcal{T}= \emptyset, c=a \} .
	\end{multline}
Each cache has $\binom{C-1}{t-1}$ shares of each file. Each user has access to $r$ caches. However, users may find the same share of a file in more than one cache. So, the effective number of shares of each file available to each user is,
\begin{equation}\label{eq24}
    m=\sum_{j=1}^{r} (-1)^{j+1} \binom{r}{j}\binom{C-j}{t-j}.
\end{equation}
The generation of the vectors $\{D^{a}_{\mathcal{G},\mathcal{T}}: a\in \mathcal{G} \}$ is done in such a way that all of them are required to reconstruct $D_{\mathcal{G},\mathcal{T}}$. For each $c\in[C]$, there are $\binom{C-1}{r-1}$ number of $r$ sized subsets of $C$ denoted by $\mathcal{G}$ such that $c\in \mathcal{G}$. For each $\mathcal{G} \in \Omega_{r}$, there are $\binom{C-r}{t}$ number of $t$ sized subsets of $c$ such that $\mathcal{G} \cap \mathcal{T} = \emptyset$. \\
    \textit{Cache Memory}: By the placement given in (\ref{eq23}), the size of each cache is 
    \begin{equation} \label{eq25}
        M= \frac{N\binom{C-1}{t-1}}{\binom{C-r}{t}}+\binom{C-1}{r-1}.
    \end{equation}
\textit{Delivery Phase}
During this phase, the user $U_{\mathcal{G}}, \mathcal{G} \in \Omega_{r}$ demands $W_{{d}_{\mathcal{G}}}$, for some ${d}_{\mathcal{G}} \in [N] $. In order to satisfy the demands, the server transmits $X=[T_{\mathcal{S}}, \textbf{q}_{\mathcal{G}}] \quad \forall \mathcal{S} \in \Omega_{t+r}, \mathcal{G} \in \Omega_{r}  $, where
\begin{equation} \label{eq26}
		{	\textbf{q}_{\mathcal{G}} = \textbf{e}_{{d}_{{\mathcal{G}}}} \oplus \textbf{p}_{G}, \quad \forall \mathcal{G} \in \Omega_{r} },
	\end{equation}
	\begin{equation} \label{eq27}
		{T_{\mathcal{S}} = V_{\mathcal{S}} \oplus {\underset{\substack{\mathcal{G}\subset \mathcal{S}\\|\mathcal{G}|=r}}{\bigoplus} \tilde{W}_{\textbf{q}_{\mathcal{G}},\mathcal{S} \backslash \mathcal{G}}} \quad \forall \mathcal{S} \in \Omega_{t+r}}.
	\end{equation}
\textit{Correctness}: Each user should get its demanded file by the above delivery. Consider an user $U_{\mathcal{H}} , \mathcal{H} \subset \mathcal{S}, \mathcal{S} \in \Omega_{t+r}.$ The transmission $T_{\mathcal{S}}$ can be written as  
	\begin{equation*}
		T_{\mathcal{S}} = V_{\mathcal{S}} \oplus \tilde{W}_{{{d}_{\mathcal{H}}},\mathcal{S}\backslash \mathcal{H}} \oplus \tilde{W}_{\textbf{p}_{\mathcal{H}},\mathcal{S}\backslash \mathcal{H}} {\underset{\substack{\mathcal{G}\subset \mathcal{S}\\|\mathcal{G}|=r \\ \mathcal{G}\neq \mathcal{H}} }{\bigoplus}  \tilde{W}_{\textbf{q}_{\mathcal{G}},\mathcal{S}\backslash \mathcal{G}}}.
	\end{equation*}
 The user $U_{\mathcal{H}}$ can get $D_{\mathcal{H},\mathcal{S} \backslash \mathcal{H}} = \tilde{W}_{\textbf{p}_{\mathcal{H}},\mathcal{S}\backslash \mathcal{H}} \oplus V_{\mathcal{S}}$ from the content of the caches accessible to it as
 \begin{equation} \label{eq28}
     D_{\mathcal{H},\mathcal{S} \backslash \mathcal{H}}=  \bigoplus_{a \in \mathcal{H}} D^{a}_{\mathcal{G}, \mathcal{S} \backslash \mathcal{H}}.  
\end{equation}
 Now, consider the term 
		 $\tilde{W}_{\textbf{q}_\mathcal{G},\mathcal{S}\backslash \mathcal{G}}$,
	
 The coefficients of $\tilde{W}_{i,\mathcal{S}\backslash \mathcal{G}},  \forall i \in [N]$ are known to the user as $\textbf{q}_{\mathcal{G}} , \forall \mathcal{G} \in \Omega_{r} $ are sent by the server and the user $U_{\mathcal{H}}$ has access to the shares $\tilde{W}_{i,\mathcal{S}\backslash \mathcal{G}}, \mathcal{G} \subset \mathcal{S},|\mathcal{G}|=r, \mathcal{G}\neq \mathcal{H} \forall i \in [N]$. So it can calculate the entire term. Thus the user will be able to get $\tilde{W}_{{d}_{{\mathcal{H}}},\mathcal{S}\backslash \mathcal{H}}$. Similarly, from all the transmissions corresponding to such $t+r$ sized subsets $\mathcal{S}$, with $\mathcal{H}\subset \mathcal{S}$, user $U_{\mathcal{H}}$ gets the missing shares of its demand. Thus, from server transmissions, the user $U_{\mathcal{H}}$ gets $\binom{C-r}{t}$ shares of the demanded file. From the caches that are accessible to it, it can get $m$ shares. Thus, it gets $\binom{C-r}{t}+m=\binom{C}{t}$ shares of the demanded file, which are enough to get back its demanded file. Similarly, any user $U_{\mathcal{G}},  \mathcal{G}\subset C,|\mathcal{G}|=r$ can recover the missing shares of its demanded file.  \\
\textit{Proof of Secrecy}: No user must gain any information about any file using only the contents of the caches that are accessible to it. By using the contents of caches and transmitted signals by the server, no user should gain any information about the files it didn't request.
To prove secrecy, first, let's show that a user cannot gain any information about any file using it's cache contents alone. We make use of the procedure given in \cite{MaS} to prove this. As each user has access to $r$ caches, the amount of cache memory available is $rM$. But, from the placement, it is clear that some shares are common in the caches that are accessible to it. So, the effective number of shares of each file that are accessible to each user in this scheme is $m=\sum_{j=1}^{r} (-1)^{j+1} \binom{r}{j}\binom{C-j}{t-j}$. The shares of all files are generated in the same way. Consider any $m$ shares of the file $W_{i}, i\in [N]$. They were generated as follows:  
\[
\begin{bmatrix}
    \tilde{W}_{i,\mathcal{T}_{1}} \\
    \tilde{W}_{i,\mathcal{T}_{2}} \\
    \vdots \\
    \tilde{W}_{i,\mathcal{T}_{m}}
\end{bmatrix} 
=
\begin{bmatrix}
    c_{1,1} &c_{1,2} &\cdots &c_{1,\binom{C}{t}} \\
    c_{2,1} &c_{2,2} &\cdots &c_{2,\binom{C}{t}} \\
    \vdots &\vdots &\ddots &\vdots \\
    
    c_{m,1} &c_{m,2} &\cdots & c_{m,\binom{C}{t}}
    
\end{bmatrix}
\cdot
\begin{bmatrix}
    W_{i,1} \\
    . \\
    W_{i,\binom{C-r}{t}} \\
    Y_{i,1} \\
    . \\
    Y{i,m}
\end{bmatrix}
\]
Let $\textbf{C}_{1}$ and $\textbf{C}_{2}$ be two sub-matrices of $\textbf{C}$ of size $m\times (n-m)$ and $m\times m$ respectively. Then, the above equation can be written as
\[
\begin{bmatrix}
    \tilde{W}_{i,\mathcal{T}_{1}} \\
    \tilde{W}_{i,\mathcal{T}_{2}} \\
    \vdots \\
    \tilde{W}_{i,\mathcal{T}_{m}}
\end{bmatrix} 
=
\textbf{C}_{1}   \begin{bmatrix}
    W_{i,1} \\
    W_{i,2} \\
    \vdots \\
    W_{i,\binom{C-r}{t}} \\
    
\end{bmatrix} + \textbf{C}_{2}
\begin{bmatrix}
    Y_{i,1} \\
    Y_{i,1} \\
    \vdots \\
    Y{i,m}
\end{bmatrix}
\]

For the collection to leak the information, there should be a non-zero matrix $\textbf{H}$ such that
\begin{equation*}
    \textbf{H}\textbf{C}_{1} \neq \textbf{0} ,\hspace{1cm}  \textbf{H}\textbf{C}_{2} = \textbf{0},   
\end{equation*}
where \textbf{0} is the zero matrix. However, the submatrix of a Cauchy matrix is a full-rank matrix. The rows of $\textbf{C}_{2}$ are linearly independent, which implies the non-existence of such \textbf{H}. Thus, information cannot be leaked from the shares of files available to each user from its cache. Now, consider the transmissions sent by the server. Each transmission is protected by a key that is available only to the user for which that particular transmission is useful as $D_{\mathcal{G},\mathcal{T}}$ is recoverable only when all $\{D^{a}_{\mathcal{G},\mathcal{T}}: a \in \mathcal{G}\}$ are available. The user $U_{\mathcal{G}}$ knows only $D_{\mathcal{G},\mathcal{T}}$, but not $V_{\mathcal{G} \cup \mathcal{T}}$.  Thus, the scheme achieves secrecy. \\
\textit{Privacy}: We now prove the condition for privacy. Each user $U_{\mathcal{G}}, \mathcal{G} \in \Omega_{r}$ has access to $r$  caches and $Z_{\mathcal{G}}, \mathcal{G} \in \Omega_{r}$ denote the content of the caches accessible to the user $U_{\mathcal{G}}$. We have
 \begin{subequations}
		\begin{align} \label{eq29}
		&I(\textbf{d}_{\bar{\mathcal{G}}}; X, \textbf{d}_{\mathcal{G}}, Z_{\mathcal{G}})\\
		&\leq I(\textbf{d}_{\bar{\mathcal{G}}}; X, \textbf{d}_{\mathcal{G}}, Z_{\mathcal{G}},W_{[N]},Y_{[N],[m]}) \\
		&=I(\textbf{d}_{\bar{\mathcal{G}}}; \textbf{q}_{\Omega_{r}}, \{{T}_{\mathcal{S}}\}_{\mathcal{S} \in \Omega_{t+r}}, \textbf{d}_{\mathcal{G}}, Z_{\mathcal{G}},W_{[N]},Y_{[N],[m]} )\\
		& \leq I(\textbf{d}_{\bar{\mathcal{G}}}; V_{\Omega_{t+r}},\textbf{q}_{\Omega_{r}}, \textbf{d}_{\mathcal{G}}, Z_{\mathcal{G}},W_{[N]},Y_{[N],[m]} ) \label{eq29d} \\ 
		&=0, \label{eq29e}
		\end{align}
	\end{subequations}
 where (\ref{eq29d}) comes from the fact that $\{{T}_{\mathcal{S}}\}_{\mathcal{S} \in \Omega_{t+r}}$ is determined by $\textbf{q}_{{\Omega_{r}}} , V_{\Omega_{t+r}} , W_{[N]},Y_{[N],[m]}$ and  (\ref{eq29e}) comes from the fact that $\textbf{d}_{\bar{\mathcal{G}}} = \textbf{q}_{\bar{\mathcal{G}}} \oplus \textbf{p}_{\bar{\mathcal{G}}}$ is independent of $V_{ \Omega_{t+r}},\textbf{q}_{\Omega_{r}}, \textbf{d}_{\mathcal{G}}, Z_{\mathcal{G}},W_{[N]}$, and $Y_{[N],[m]}$ as $\textbf{p}_{\bar{\mathcal{G}}}$ is independently and uniformly distributed over $\mathcal{F}_{2}^{N}$.
 
 This completes the proof of Theorem \ref{th2}.
\subsection{Proof of Theorem \ref{th3}}\label{ssec5}
In this sub-section, we obtain a secretive scheme that needs only secrecy condition to be satisfied.

In the secretive and private scheme proposed in Section \ref{ssec5a}, we get the scheme that requires only secrecy condition by making $\textbf{p}_{\mathcal{G}}=0, \forall \mathcal{G} \in \Omega_{r}$ for $t\in [0:C-r-1]$. The memory and rate equations remain the same for $t\in[0:C-r-1]$. When $t=C-r$, the number of transmissions done by the server is one, which means all the users require that transmission. So, there is no need to protect the transmission from any user. The files are encoded using a $(K-1,K)$ non-perfect secret sharing scheme\cite{CDN}. In that case, the cache placement is as follows:
\begin{equation} \label{eq30}
Z_{c} = \{\tilde{W}_{i,\mathcal{T}} : c\in\mathcal{T} , \mathcal{T} \in \Omega_{C-r} \quad \forall \quad i \in [N] \}.
\end{equation}
By the above placement, the size of each cache when $t=C-r$ is
\begin{equation} \label{eq31}
    M= \frac{N\binom{C-1}{t-1}}{\binom{C-r}{t}}=N\binom{C-1}{r}.
\end{equation}
During the delivery phase, the server transmits
\begin{equation*}
T =  {\underset{\substack{\mathcal{G}\subset [C]\\|\mathcal{G}|=r}}{\bigoplus} \tilde{W}_{d_{\mathcal{G}},[C]\backslash \mathcal{G}}}.
\end{equation*}
Consider any user $U_{\mathcal{H}}$,$\mathcal{H} \in \Omega_{r}$. It is clear that it can get its required share $\tilde{W}_{d_{\mathcal{H}}, [C]\backslash \mathcal{H}}$ from the transmission. Similarly, all the users will be able to get the files they demand, and they know nothing about the files they didn't request because each user can access only $K-1$ shares of all the files from the caches accessible to them. The achieved rate is $1$.

This completes the proof of Theorem \ref{th3}.  
\begin{exmp}
    Consider $C=4,r=2,N=6$. Each file is divided into $2$ subfiles. Thus, $W_{i}=\{W_{i,1},W_{i,2}\} \quad \forall i\in[3]$. Number of users, $K=\binom{C}{r}=\binom{4}{2}=6$. The $6$ users are $\{U_{\{1,2\}}, U_{\{1,3\}}, U_{\{1,4\}}, U_{\{2,3\}}, U_{\{2,4\}}, U_{\{3,4\}}\}$. For every file, $4$ shares are generated using $(2,4)$ non-perfect secret sharing scheme. To generate shares of a file $W_{i}$, first form a $4 \times 1$ column vector comprised of $2$ subfiles and $2$ independent random vectors $\{ Y_{i,1}, Y_{i,2} \}$, each uniformly distributed over $\mathcal{F}_{2}^{F/2}$. We consider a $4\times 4 $ cauchy matrix over $\mathcal{F}_{2^{3}}$. For each file $W_{i}, i \in [6]$, the shares are obtained as follows:
    \[
\begin{bmatrix}
    \tilde{W}_{i,{1}} \\
    \tilde{W}_{i,{2}} \\
    \tilde{W}_{i,{3}} \\
    \tilde{W}_{i,{4}} \\
    
\end{bmatrix} 
=
\begin{bmatrix}
    1 &6 &2  &4  \\
    6 &1 &4  &2  \\
    2 &4 &1  &6  \\
    4 &2 &1  &6  \\
    
\end{bmatrix}
\cdot
\begin{bmatrix}
    W_{i,1} \\
    W_{i,2} \\
    
    Y_{i,1} \\
    Y{i,2}
\end{bmatrix}
\]

    The server generates random vectors $\{V_{\{1,2,3\}},V_{\{1,2,4\}}, V_{\{1,3,4\}}, V_{\{2,3,4\}}\}$, each independently and uniformly from $\mathcal{F}_{2}^{F/2}$. Now, the server generates $K=6$ random vectors as follows:
	\begin{center}
		$\textbf{p}_{\mathcal{G}} \triangleq (p_{\mathcal{G},1},p_{\mathcal{G},2},..,,p_{\mathcal{G},6})^{T} \sim Unif\{\mathcal{F}_{2}^{3}\}, \forall \mathcal{G}\in \Omega_{2}$.
	\end{center}
    The  privacy keys , denoted by ${\{\tilde{W}_{\textbf{p}_{\mathcal{G}},\mathcal{T}}: \mathcal{G} \in \Omega_{2} , \mathcal{T} \in [4]}$, \\ ${\mathcal{G} \cap \mathcal{T}= \emptyset\}}$ are generated as :
		\begin{center}
			${\tilde{W}_{\textbf{p}_{\mathcal{G}},\mathcal{T}}\triangleq \underset{n\in[6]}{\bigoplus} p_{\mathcal{G},n}.\tilde{W}_{n,\mathcal{T}}}$.
		\end{center}
		Let
		\begin{center}
			$D_{\mathcal{G},\mathcal{T}} \triangleq \tilde{W}_{\textbf{p}_{\mathcal{G}},\mathcal{T}} \oplus V_{\mathcal{G} \cup \mathcal{T}} ,\quad  \mathcal{G} \in \Omega_{2} , \mathcal{T} \in [4],\mathcal{G} \cap \mathcal{T}= \emptyset $.
		\end{center}
	For every $D_{\mathcal{G},\mathcal{T}}$, the following is done:
\begin{itemize}
    \item  For any $a\in\mathcal{G}$, generate a random vector $D^{a}_{\mathcal{G},\mathcal{T}}$  independently and uniformly from $\mathcal{F}_{2}^{F/2}$.
    \item Now, generate $D^{\mathcal{G}\backslash a}_{\mathcal{G},\mathcal{T}}$, as follows 
    \begin{equation}
        D^{\mathcal{G}\backslash a}_{\mathcal{G},\mathcal{T}} =  D^{a}_{\mathcal{G},\mathcal{T}}  \oplus D_{\mathcal{G},\mathcal{T}}.
    \end{equation}
\end{itemize}
    Now, let's look at the placement. The caches are filled as follows: \\
    $Z_{1} = \{\tilde{W}_{{1,1}}, \tilde{W}_{{2,1}}, \tilde{W}_{{3,1}}, \tilde{W}_{{4,1}}, \tilde{W}_{{5,1}}, \tilde{W}_{{6,1}}\} \cup \\ \{ D^{1}_{\{1,2\},3}, D^{1}_{\{1,2\},4}, D^{1}_{\{1,3\},2}, D^{1}_{\{1,3\},4}, D^{1}_{\{1,4\},3}, D^{1}_{\{1,4\},2} \},$ \\
		 $Z_{2} = \{\tilde{W}_{{1,2}}, \tilde{W}_{{2,2}}, \tilde{W}_{{3,2}}, \tilde{W}_{{4,2}}, \tilde{W}_{{5,2}}, \tilde{W}_{{6,2}}\} \cup \\ \{ D^{2}_{\{1,2\},3}, D^{2}_{\{1,2\},4}, D^{2}_{\{2,3\},1}, D^{2}_{\{2,3\},4}, D^{2}_{\{2,4\},3}, D^{2}_{\{2,4\},1} \},$ \\  
		 $Z_{3} = \{\tilde{W}_{{1,3}}, \tilde{W}_{{2,3}}, \tilde{W}_{{3,3}}, \tilde{W}_{{4,3}}, \tilde{W}_{{5,3}}, \tilde{W}_{{6,3}}\}\cup \\ \{ D^{3}_{\{1,3\},2}, D^{3}_{\{1,3\},4}, D^{3}_{\{2,3\},1}, D^{3}_{\{2,3\},4}, D^{3}_{\{3,4\},2}, D^{3}_{\{3,4\},1} \}, $ \\
$Z_{4} = \{\tilde{W}_{{1,4}}, \tilde{W}_{{2,4}}, \tilde{W}_{{3,4}}, \tilde{W}_{{4,4}}, \tilde{W}_{{5,4}}, W_{{6,4}}\} \cup \\ \{ D^{4}_{\{1,4\},3}, D^{4}_{\{1,4\},2}, D^{4}_{\{2,4\},1}, D^{4}_{\{2,4\},3}, D^{4}_{\{3,4\},2}, D^{4}_{\{3,4\},1} \}. $ \\
Based on the above placement, the size of each cache is $6$.
        Let $d_{\{1,2\}}=1, d_{\{1,3\}}=2, d_{\{1,4\}}=3, d_{\{2,3\}}=4, d_{\{2,4\}}=5, d_{\{3,4\}}=6$. The server transmissions are as follows:\\
        $T_{\{1,2,3\}}= V_{\{1,2,3\}} \oplus \tilde{W}_{\textbf{q}_{\{1,2\}},3} \oplus \tilde{W}_{\textbf{q}_{\{1,3\}},2} \oplus \tilde{W}_{\textbf{q}_{\{2,3\}},1} ,$ \\
        $T_{\{1,2,4\}}= V_{\{1,2,4\}} \oplus \tilde{W}_{\textbf{q}_{\{1,2\}},4} \oplus \tilde{W}_{\textbf{q}_{\{1,4\}},2} \oplus \tilde{W}_{\textbf{q}_{\{2,4\}},1} ,$ \\
        $T_{\{1,3,4\}}= V_{\{1,3,4\}} \oplus \tilde{W}_{\textbf{q}_{\{1,3\}},4} \oplus \tilde{W}_{\textbf{q}_{\{1,4\}},3} \oplus \tilde{W}_{\textbf{q}_{\{3,4\}},1} $ ,\\
        $T_{\{2,3,4\}}= V_{\{2,3,4\}} \oplus \tilde{W}_{\textbf{q}_{\{2,3\}},4} \oplus \tilde{W}_{\textbf{q}_{\{2,4\}},3} \oplus \tilde{W}_{\textbf{q}_{\{3,4\}},2} $ .\\
        $\textbf{q}_{\mathcal{G}} = \textbf{p}_{\mathcal{G}} + \textbf{e}_{d_{\mathcal{G}}} \quad \forall \mathcal{G} \in \Omega_{2}. $ \\
        Now, consider the user $U_{\{1,2\}}$ and the transmission $T_{\{1,2,3\}}$. The term $\tilde{W}_{\textbf{q}_{{\{1,3\}}},2} $ can be written as  $\underset{i \in [6]}{\bigoplus} {q}_{{\{1,3\},i}} \tilde{W}_{i,2}$. As, the user knows $\textbf{q}_{\{1,3\}}=({q}_{{\{1,3\},1}},.....{q}_{{\{1,3\},6}})$ and has access to the $2^{nd}$ cache, it can calculate the term. Similarly, the user $U_{\{1,2\}}$ can calculate the term $\tilde{W}_{\textbf{q}_{{\{2,3\}}},1} $. Moreover, the user $U_{\{1,2\}}$ can get the key $V_{\{1,2,3\}} \oplus \tilde{W}_{\textbf{p}_{\{1,2\}},3} $ from the caches $1$ and $2$. The user $U_{\{1,2\}}$ can get $D_{\{1,2\},3}$  from the content of the caches accessible to it as
 \begin{equation}
     D_{\{1,2\},3}=  D^{1}_{\{1,2\},3} \oplus D^{2}_{\{1,2\},3}.
\end{equation} 
So, the user $U_{\{1,2\}}$ can decode $\tilde{W}_{{d}_{{\{1,2\}}},3}$. Similarly the user can get $\tilde{W}_{{d}_{{\{1,2\}}},4}$ from $T_{\{1,2,4\}}$. Thus, it has all $4$ shares of its demanded file. So, it can reconstruct its demanded file. Now, the transmissions $T_{\{1,3,4\}}$ and $T_{\{2,3,4\}}$ are protected by a key which prevents the user $U_{\{1,2\}}$ from getting any information about any file it didn't request. Through its caches, it has access to only $2$ shares of all the files. By the nature of $(2,4)$ non-perfect secret sharing scheme, it doesn't know anything about the files it didn't request. Although the user $U_{\{1,2\}}$ knows $D_{\{1,2\},3}$, it doesn't know $V_{\{1,2,3\}}$. So, it learns nothing about the file it didn't request. The vectors $\textbf{q}_{\mathcal{G}},  \forall \mathcal{G} \in \Omega_{2} $ are random vectors independently and uniformly distributed over $ \mathcal{F}^{6}_{2}$. So, the demand privacy is satsified.  
\end{exmp}
  \section{Numerical Evaluations} \label{sec6}
  \begin{figure} 
     \centering
     \includegraphics[width=\linewidth]{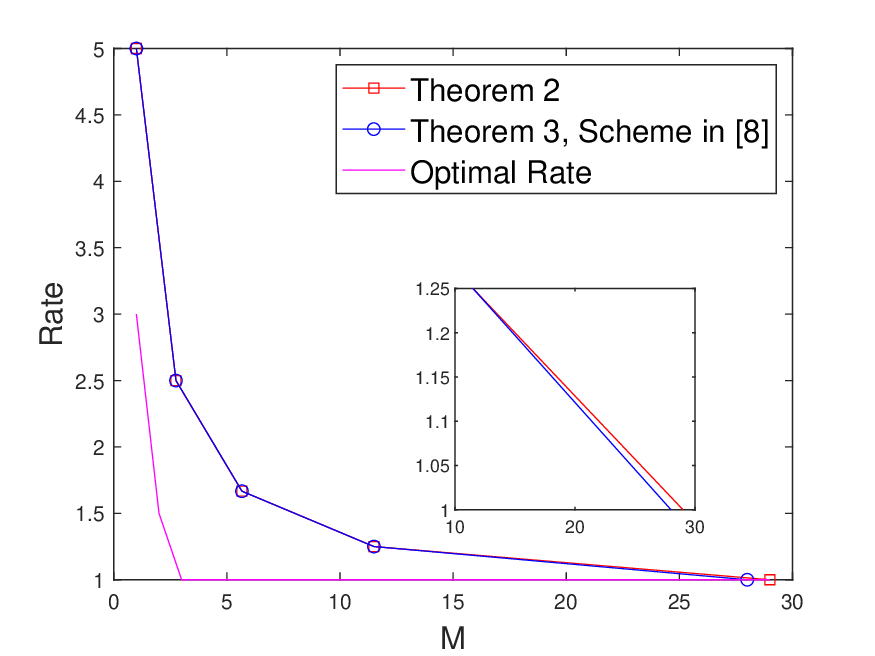}
     \caption{Performance comparison of the proposed schemes with \cite{RPKP} when C=5, r=1, N=7}
     \label{Fig2}
\end{figure}
\begin{figure} 
     \centering
     \includegraphics[width=\linewidth]{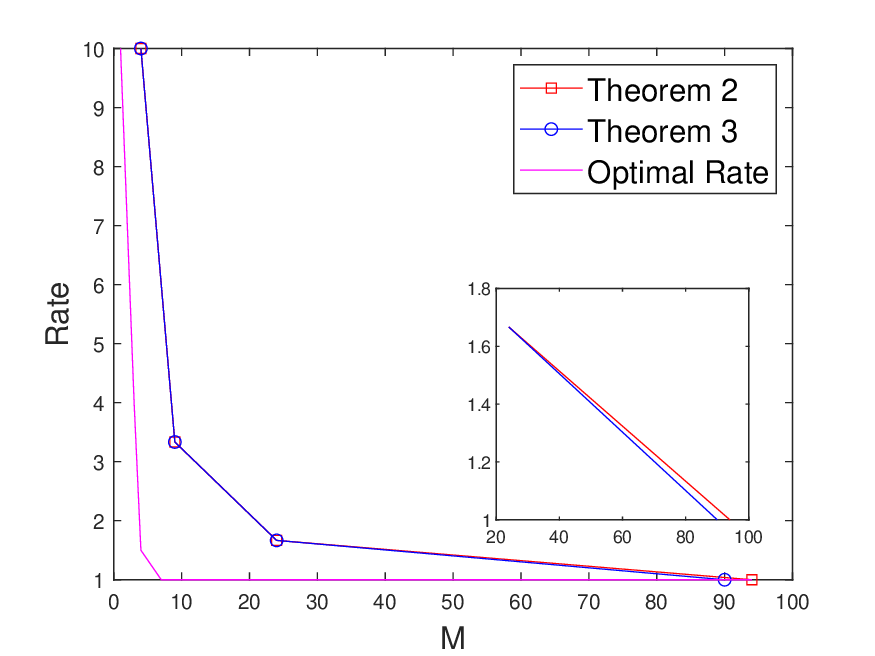}
     \caption{Performance comparison of the proposed schemes when C=5, r=2, N=15.}
     \label{Fig3}
\end{figure}
\begin{figure} 
     \centering
     \includegraphics[width=\linewidth]{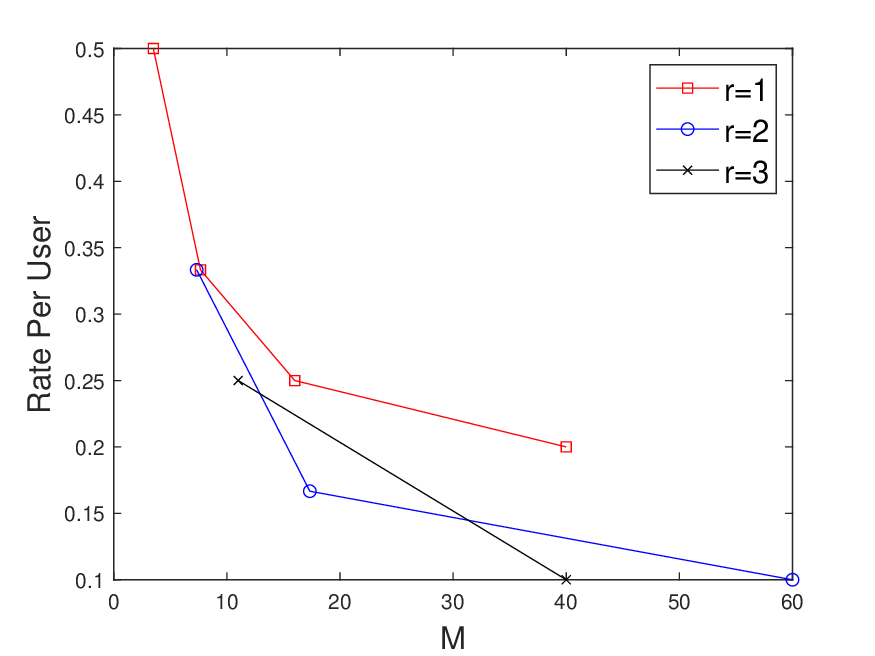}
     \caption{Performance comparison with \cite{RPKP} for different values of $r$ considering the same cache size and the same number of caches when C=5, N=10.}
     \label{Fig4}
\end{figure}
\begin{figure} 
     \centering
     \includegraphics[width=\linewidth]{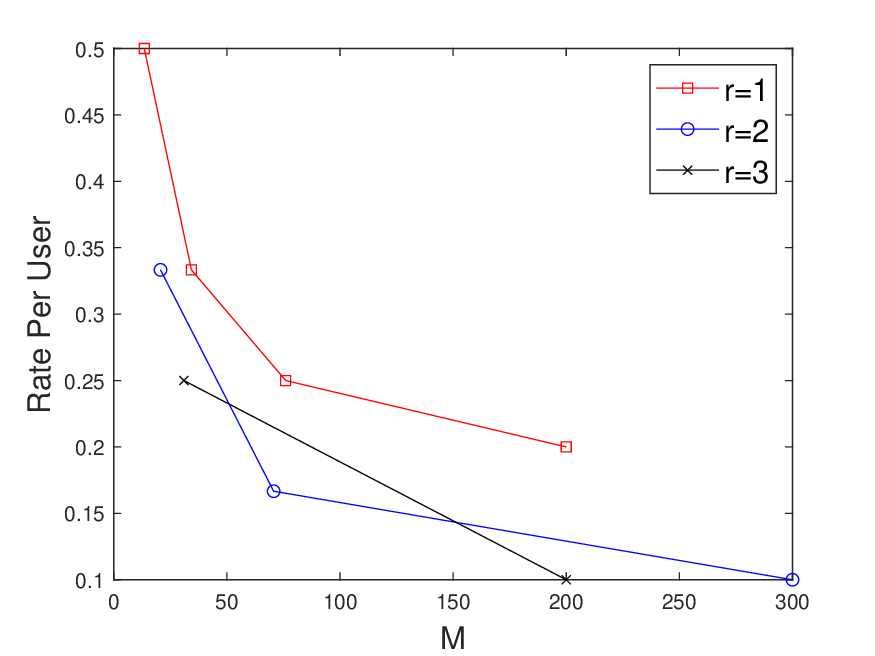}
     \caption{Performance comparison with \cite{RPKP} for different values of $r$ considering the same cache size and the same number of caches when C=5, N=50.}
     \label{Fig5}
\end{figure}
\begin{figure} 
     \centering
     \includegraphics[width=\linewidth]{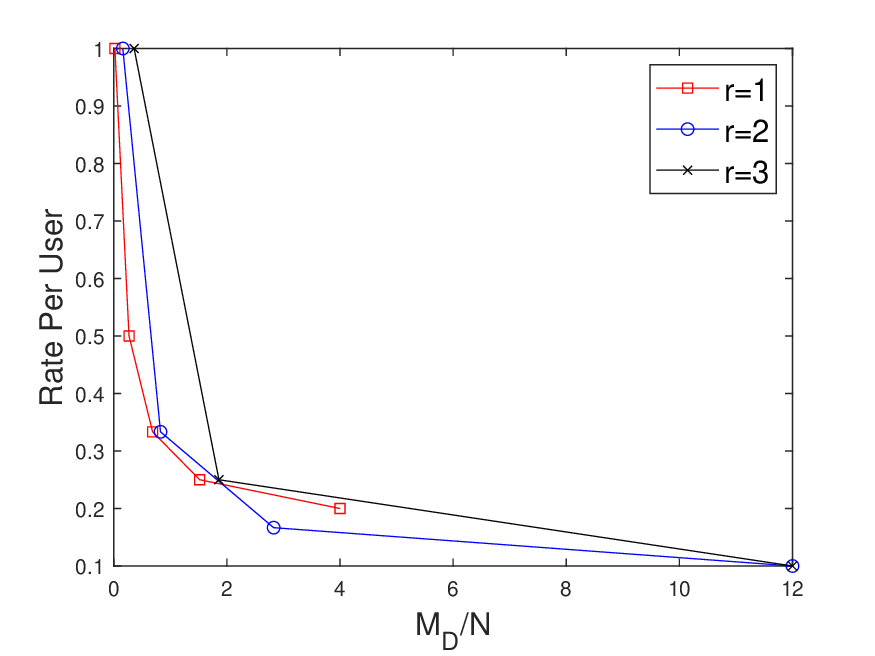}
     \caption{Performance comparison with \cite{RPKP} for different values of $r$ considering the same total memory accessed by each user when C=5, N=50.}
     \label{Fig7}
\end{figure}
\begin{figure} 
     \centering
     \includegraphics[width=\linewidth]{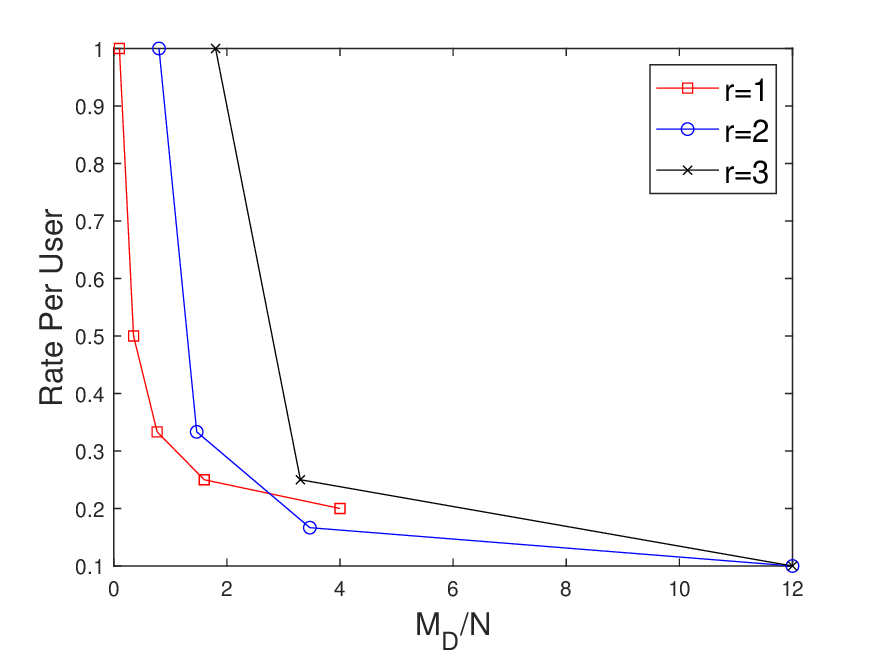}
     \caption{Performance comparison with \cite{RPKP} for different values of $r$ considering the same total memory accessed by each user when C=5, N=10.}
     \label{Fig8}
\end{figure}

  In this section, we analyze the schemes proposed in this work numerically.
\subsection{Comparison of Theorem \ref{th2} and Theorem \ref{th3} with \cite{RPKP}}
In this sub-section, we numerically compare Theorem \ref{th2} and Theorem \ref{th3} with the scheme for dedicated cache setup given in \cite{RPKP}. The scheme that provides both secrecy and privacy for combinatorial topology is given in Theorem \ref{th2}. The scheme that provides only secrecy for combinatorial topology is given in Theorem \ref{th3}. When $r=1$, the setup considered in this work reduces to the dedicated cache setup in \cite{MaN}. Consider Fig.\ref{Fig2}. It shows the comparison of Theorem\ref{th2}, and Theorem \ref{th3} with the scheme in \cite{RPKP} when $C=5, r=1, N=7$. The proposed scheme in Theorem \ref{th2} provides both secrecy and privacy for the same memory and rate given by the scheme in \cite{RPKP} except from $t= C-2$ to $t= C-1 $. So, from $t= C-2$ to $t= C-1 $ there is a gap between Theorem \ref{th2} and Theorem \ref{th3} curves in Fig.\ref{Fig2}. At this point, providing both secrecy and privacy requires more memory than providing only secrecy. In Fig.\ref{Fig2}, it is shown exclusively from cache size, $M=11.5$ to $M=29$.  Theorem \ref{th3} reduces to the scheme in \cite{RPKP} when $r=1$. From Fig.\ref{Fig2}, it is clear that Theorem \ref{th2} and Theorem \ref{th3} always coincide except for a very small memory region. Even in that region, the difference is very small. Fig. \ref{Fig3} shows the plots for Theorem \ref{th2} and Theorem \ref{th3} when $C=5,r=2$. The secretive scheme in Theorem \ref{th3} and the secretive and private scheme in Theorem \ref{th2} are plotted along with the optimal rate given by the lower bound in Theorem \ref{th1}. Having both secrecy and privacy needs additional memory only from $t=2$ to $t =3$. It is shown exclusively in Fig.\ref{Fig3} from $M=24$ to $M=94$.
\subsection{Comparison of Theorem \ref{th3} with \cite{RPKP}}
In this sub-section, we compare the proposed scheme in Theorem \ref{th3} with the scheme for dedicated cache setup given in \cite{RPKP}. The scheme in \cite{RPKP} considers secrecy for dedicated cache setup, whereas the scheme in Theorem \ref{th3} considers secrecy for the multi-access setup. The comparison  is done in the following ways:
  \begin{itemize}
      \item Firstly, we consider the total number of caches in both the dedicated cache and multi-access setup to be the same. The size of the cache is also the same. But, the multi-access setup supports more number of users ($K=\binom{C}{r}$) than the dedicated cache setup for a given $r$. So, we compare the Rate Per User (RPU), which is the rate normalized with the number of users instead of the rate. The total system memory is also the same in both the dedicated cache and the multi-access setup.
      \item Then we consider the case where the total number of caches is the same, and each user accesses the same amount of memory. To make the total memory accessed per user the same for both dedicated cache and multi-access settings, we make the cache size of the dedicated cache setup $r$ times more than the cache size of the multi-access setup.
       
  \end{itemize}
  Now, we discuss the numerical evaluations considering the above two settings. We use the subscript $D$ and $M$ for dedicated cache and multi-access cache setups, respectively. \\
  1) \textit{Same number of cache and same cache size }: 
  In a dedicated cache setup, the number of users and the number of caches is the same. So, $C_{D}=K_{D}$. But, in a multi-access cache setup, the number of users is more than the number of caches as $K_{M}=\binom{C_{M}}{r}$. We consider the number of caches to be the same in both setups. So, $C_{M}=C_{D}=K_{D}=C$. To be able to compare two setups with different number of users, we use the rate per user. When $r=1$, the curve for the secretive scheme in Theorem \ref{th3} coincides with the secretive scheme for the dedicated cache setup in \cite{RPKP}. So, we use the $r=1$ curve to refer to the performance of the scheme in \cite{RPKP}. For $C=5, N=10$, the curves for different values of $r$ is shown in Fig.\ref{Fig4}. When $t=0$, to achieve the same per-user rate of $1$, the dedicated cache setup requires cache memory size of $M=1$, whereas the multi-access setup requires $M=\binom{C-1}{r-1}$. So, at that point, the dedicated cache setup performs better than the multi-access setup. The curves for other values of $t$ are shown in Fig.\ref{Fig4} when $C=5, N=10$. Clearly, multi-access curves when $r=2$ and $r=3$ are below the curve for dedicated cache setup, indicating better performance. But, among the multi-access setups with $r=2$ and $r=3$, the curve for $r=3$ performs better in the higher memory region when $M\geq31.5$. The performance of a multi-access setup depends on both the value of $r$ and the memory size $M$. In order to provide secrecy, the shares are generated for each file. As the value of $r$ increases, the size of each share increases. So, a multi-access setup with more $r$ needs more cache memory to store a single share. The amount of memory required to store keys is also more in a multi-access setup. These are disadvantages in terms of performance. But, if the value of $r$ is higher, the total number of transmissions sent by the server is less as the amount of memory accessed by each user is greater. So, having more $r$ has both advantages and disadvantages. The overall performance depends on the combined effect of these two factors. Fig.\ref{Fig5} shows the plots when $C=5, N=50$. From Fig.\ref{Fig4}, Fig.\ref{Fig5}, the curve with $r=3$ has the advantage in the higher memory region when compared to the curve with $r=2$.   \\
  2) \textit{Same number of caches and total memory accessed per user}: 
  Now, we consider that the total number of caches in the dedicated cache setup and the multi-access setup are the same. We compare the performance in terms of total memory accessed per user. Each user in a multi-access setup accesses the memory of size $r*M_M$. But, in the dedicated cache, each user accesses the memory of size $M_{D}$. So, to have the same total memory accessed per user, $M_{D}=r*M_{M}$. In this case, the total system memory of the dedicated cache setup is also $r$ times more than that of the multi-access setup. The plots are shown in Fig.\ref{Fig7} and Fig.\ref{Fig8} for $C=5, N=50$ and $C=5,N=10$ respectively. The performance of the dedicated cache is better than that of the multi-access setup. In a multi-access setup, the same share of a file is present in more than one cache, whereas in a dedicated cache setup, each user accesses one cache, and the no share is repeated. So, effectively each user in the dedicated cache setup can access more shares. The amount of memory required to store keys is also higher in the multi-access setup. In a multi-access setup, the size of each share of a file is larger than that of a dedicated cache setup. As total memory is also more for dedicated cache setup, multi-access setup is at a disadvantage in this comparison.
  \section{Conclusion}\label{sec7}
  In this work, we considered secrecy and demand privacy in combinatorial topology. We proposed a scheme that provides both secrecy and privacy. We derived a lower bound based on cut-set arguments and proved that our scheme is optimal when $r\geq C-1$ and order-optimal when $M$ is greater than a certain value for $r<C-1$. In most of the achievable memory region, for the same rate given by the secretive scheme in \cite{RPKP}, our scheme provides both secrecy and privacy when $r=1$. We numerically compared our scheme with the scheme for the dedicated cache setup in \cite{RPKP}. 
\section*{Acknowledgment}
	This work was supported partly by the Science and Engineering Research Board (SERB) of Department of Science and Technology (DST), Government of India, through J.C. Bose National Fellowship to B. Sundar Rajan.
  \appendix 
  \begin{center}
		\bf{Proof of Theorem 4}
	\end{center}

  Consider the problem setup given in Section \ref{sec2}. The server has $N$ files, each of size $F$ bits, and it is connected to $K$ users through an error-free shared link. Each user has access to a unique set of $r$ out of $C$ caches. For every set of $r$ caches, there is a user. The rate of the scheme with $M= \frac{N\binom{C-1}{t-1}}{\binom{C-r}{t}}+\binom{C-1}{r-1} $ that satisfies the decodability and secrecy  condition is given by,
    \begin{equation} \label{eq34}
    R( M)= \frac{\binom{C}{t+r}}{\binom{C-r}{t}}.
    \end{equation}
    The optimal rate for $N\geq 2K$ is given by,
    \begin{equation*}
      R^{*} \geq \max_{l \in \{1,2,...K\}} \frac{l \lfloor N/l \rfloor-1-((z-r) M)}{\lfloor N/l \rfloor - 1},
  \end{equation*}
  where $z=\min(l+r-1,C)$.
   
  We prove the optimality results using the above inequality. Let's consider the case when $r=C-1$. The possible values of $t$ are $0$ and $1$. When $t=0$, the size of cache is
  \begin{equation}
      M=\binom{C-1}{r-1}=\frac{rK}{C}=r.
  \end{equation}
  Now, consider the optimal rate inequality for $l=K$. The value of $z= \min(l+r-1,C)$ is $C$. By replacing $l$ with $k$ and $z$ with $C$ in the optimal rate inequality, we get
  \begin{subequations}
   \begin{align*}
      R^{*} 
      & \geq \max_{l \in \{1,2,...K\}} \frac{l (\lfloor N/l \rfloor)-1-((z-r) M)}{\lfloor N/l \rfloor - 1}. \\
      &=\frac{K (\lfloor N/K \rfloor)-1-((C-r) M)}{\lfloor N/K \rfloor - 1} \\
      & =\frac{K (\lfloor N/K \rfloor)-1-((C-r)r)}{\lfloor N/K \rfloor - 1} \\
    & =\frac{K (\lfloor N/K \rfloor)-(1+r)}{\lfloor N/K \rfloor - 1} \\
      & =\frac{K (\lfloor N/K \rfloor)-(K)}{\lfloor N/K \rfloor - 1} \\
      &=K.
      \end{align*}
  \end{subequations}
        For $t=0$, the achievable rate is given by 
\begin{equation} \label{eq37}
    R(rK/C)= K.
\end{equation}
Thus, optimality is proved for $t=0$. Now consider the case when $t=1, r=C-1$. The cache memory size $M$ is
\begin{equation*}
    M=\frac{N}{C-r}+\binom{C-1}{r-1}.
\end{equation*}
  Now, let's calculate the rate when $t=1,r=C-1$. The rate for the proposed scheme is given by       
      \begin{equation} \label{eq49}
    R(N/(C-r)+\binom{C-1}{r-1})= 1.
\end{equation}  
      Consider the optimal rate inequality when $l=1$.  
     \begin{subequations}
   \begin{align*}
      R^{*} 
      & \geq \max_{l \in \{1,2,...\min(N/2,K)\}} \frac{l (\lfloor N/l \rfloor)-1-((z-r) M_{M})}{\lfloor N/l \rfloor - 1}. \\
      &=\frac{(N-1)-(r-r)M}{N-1} \\
      &=1.
      \end{align*}
      \end{subequations}
        So, optimality proved for $r=C-1$. When $r=C$, there is only one user, and only one memory point at $t=0$ exists. The rate at that point is $1$, which is optimal.\\
        
        Now, consider the case when $t=1,r <C-1$. Let the memory size at $t=1$ be $M_1$. So, 
        \begin{equation*}
            M_1=\frac{N}{C-r}+\binom{C-1}{r-1}
        \end{equation*}
        The rate, $R$ for $t=1$ is given by :
\begin{equation*}
    R= \frac{\binom{C}{r+1}}{C-r}.
\end{equation*}

The above equation can be simplified as follows.
\begin{align*}
    R &= \frac{\binom{C}{r+1}}{C-r} \\
    &= \frac{C!}{(C-r-1)! (r+1)! (C-r)} \\
    &= \frac{ \binom{C}{r}}{(r+1) } .
\end{align*}
As $R^{*} \geq1$, at $M=M_1$, 
\begin{align*}
    \frac{R}{R^{*}} 
    & \leq \frac{\binom{C}{r}}{r+1} \\
    &\leq 5, \quad\text{if} \quad K\leq 5(r+1).\\
\end{align*}
Thus,
\begin{align}
    \frac{R}{R^{*}} \leq 5.
\end{align}
Now, consider any $M>M_1$. The rate at that particular $M$ cannot be greater than the rate at $M=M_1$. So, for any $M\geq M_1$, we can say that
\begin{align}
    \frac{R}{R^{*}} \leq 5.
\end{align}

This completes the proof of Theorem \ref{th4}.

\end{document}